\documentclass[aps,prb,preprint]{revtex4-1} 

\usepackage{amsthm,amsfonts,graphicx,epsfig,booktabs}

\newcommand{\rmi}{{\rm i}}

\newcommand{\rmd}{{\rm d}}
\newcommand{\omegap}{\omega_{\rm p}}
\newcommand{\omegat}{\omega_{\rm t}}

\begin{document}

\title{Gravitational dispersion in a torsional wave machine}

\author{Rafael de la Madrid}
\email{Corresponding author's e-mail: rafael.delamadrid@lamar.edu}

\author{Alejandro Gonzalez} 
\author{George M.~Irwin}

\affiliation{Department of Physics, Lamar University, Beaumont, Texas 77710}

\date{August 20, 2014}

\begin{abstract}
\noindent We demonstrate that mechanical waves traveling in a torsional, 
mechanical wave machine exhibit dispersion due to gravity and the 
discreteness of the medium. We also show that although the dispersion due 
to discreteness is negligible, the dispersion due to gravity can be easily 
measured, and can be shown to disappear in a zero-gravity environment.
\end{abstract}

\maketitle

\section{Introduction}
\label{sec:intro}

Shive's torsional wave machine\cite{SHIVE} is one of the most 
widely used apparatuses for demonstrating mechanical waves in 
physics and engineering 
courses. Because most waves travel at speeds that are difficult to view, 
Shive created his wave machine to demonstrate slower 
transverse waves. Shive's machine can be used to demonstrate phenomena
such as wave propagation, reflection and transmission at a boundary 
(fixed or free), 
constructive and destructive interference, standing waves and resonances,
and impedance matching.\cite{SHIVE,BURGEL,GREEN2,PIZZO,SKELDON1,SKELDON2,GREEN1}

According to Shive, the speed of the waves traveling along a torsional 
wave machine
is a constant that depends on the stiffness of the wire and on the 
moment of inertia of the rods. However,
Burgel~\cite{BURGEL} realized that the speed of the waves traveling
through a wave machine depends on the frequency, meaning a torsional 
wave machine is dispersive. After a thorough analysis, Burgel concluded 
that the origin of the dispersion lies in the discreteness of 
the medium (in this case, the rods through which the waves propagate).

In the present paper, we re-examine Burgel's analysis and show that there are
actually two sources of dispersion in a torsional wave machine. One source
indeed originates from the discreteness of the medium, but a second, more
important source originates from a restoring gravitational torque. We show
that for practical purposes, the restoring gravitational torque
is the only source of dispersion, whereas the effect of discreteness
on the dispersion is negligible. In addition, we show that the 
gravitationally-induced dispersion is reduced in reduced gravity, and that
it disappears in a zero-gravity environment.

The remainder of the paper is organized as follows.  In Sec.~\ref{sec:theory},
we review the wave equation and the dispersion
relation of a torsional wave machine. We also obtain the dispersion relation
in the case that the discreteness of the rods is ignored and also when
the gravitational field strength goes to zero. In Sec.~\ref{sec:exparra},
we present the experimental procedure used to measure the harmonics of 
the wave machine, and in Sec.~\ref{sec:expres}, we compare
the experimental data with the theoretical predictions.
Section~\ref{sec:conclus} contains our 
conclusions.

Due to the simplicity of the experiments, the tractability of the theory,
and the good agreement between theory and experiment,
the results of our analysis can be 
easily adapted to an undergraduate laboratory on harmonics (normal modes), 
dispersion relations, and wave dispersion.

\section{Theory}
\label{sec:theory}

In our experiments, we use two PASCO wave machines.\cite{PASCO} These 
wave machines have a central wire that is square shaped in cross section
and runs through eleven supports (see Fig.~\ref{FIG:exap}). In one of the 
wave machines, the rods are 0.228-m long (the ``short-rod machine''), whereas in the other one
the rods are 0.456-m long (the ``long-rod machine''). Each rod's center 
of mass is just below the wire, allowing the rods to be balanced 
horizontally in normal gravity (see Fig.~\ref{FIG:torque}). The long-rod
machine has $N=71$ rods and a length of $L=0.89$\,m, and the short-rod
machine has $N=72$ rods and a length of $L=0.90$\,m. 

The theory describing motion in a wave machine is well 
known.\cite{SHIVE,BURGEL,SKELDON1,SKELDON2} Each rod of the 
machine oscillates as a pendulum around its horizontal, equilibrium 
position. As the torsion wire twists, the rods tilt away from their stable, 
equilibrium position, and two restoring torques act on the rods, one 
stemming from gravity and another one from the torsion of the wire. Let us 
denote by $j$ the rod number, with $j=1,2, \ldots , N$, where 
$N$ is the total number of rods. When the $j$th rod
is tilted, its weight produces a torque about
the axis of rotation given by [see Fig.~\ref{FIG:torque}(b)]
\begin{equation}
       \tau _{\rm g}= -mgR\sin (\theta _j) \, ,
\end{equation}
where $m$ is the mass of the rod, $g$ is the gravitational field strength, 
$R$ is the rod's radius, and $\theta _j$ is the angular displacement from
the horizontal, equilibrium position.  The minus sign arises from the
restoring nature of the gravitational torque. For 
small oscillations, the gravitational torque can be approximated by
\begin{equation}
       \tau _{\rm g}= -mgR \,\theta _j \, .
           \label{gt}
\end{equation}
The torque on the $j$th rod due to the twisting of the torsion wire 
can be written as\cite{BURGEL,SKELDON1,SKELDON2}
\begin{equation}
       \tau _{\rm torsion}= \kappa _d (\theta _{j-1}-\theta _j) +
           \kappa _d (\theta _{j+1}-\theta _j) \, ,
            \label{tt}
\end{equation}
where the constant $\kappa _d$ denotes the torsion constant of a section of 
the wire of length $d$, $d$ being the distance between two consecutive 
rods. By combining Eqs.~(\ref{gt}) and~(\ref{tt}) with
Newton's second law for rotational motion, we obtain the equation of motion
\begin{equation}
  I_{\rm rod} \left( \frac{\rmd ^2 \theta _j}{\rmd t^2} \right) =
       -mgR\, \theta _j +  \kappa _d (\theta _{j-1}-\theta _j) +
           \kappa _d (\theta _{j+1}-\theta _j)  \, ,
       \label{eqmonrod}                          
\end{equation}
where $I_{\rm rod}$ denotes the moment of inertia of one 
rod. Equation~(\ref{eqmonrod}) holds for all but the end rods ($j=1$ and
$j=N$). The equation of motion of the end rods depends on the boundary
conditions (either free or fixed), as discussed in Ref.~\onlinecite{BURGEL}.

When we send a pulse through a wave machine, the shape of the pulse does not
change appreciably, and therefore the different frequencies making up the
pulse travel at nearly the same speed $c$. Thus, for most applications, one
can safely assume that the wave machine is non-dispersive. However, as
first pointed out by Burgel,\cite{BURGEL} the wave machine is actually 
dispersive. As shown in Ref.~\onlinecite{BURGEL}, the dispersion relation 
associated with Eq.~(\ref{eqmonrod}) is given by
\begin{equation}
       \omega ^2 =\omegap ^2 +4 \omegat ^2 \sin ^2 \left( \frac{kd}{2}
                                                   \right) ,
        \label{dispredtod}
\end{equation}
where
\begin{equation}
         \omegat =\sqrt{\frac{\kappa_d}{I_{\rm rod}}} \,
              \label{omegat}
\end{equation}
is the angular frequency of the torsional mode, and
\begin{equation}
         \omegap =\sqrt{\frac{mgR}{I_{\rm rod}}}
              \label{omegap}
\end{equation}
is the angular frequency of the pendulum mode, in which
the wave machine oscillates due to gravity alone
($\lambda =\infty$, $k=0$). 

In Eq.~(\ref{dispredtod}), there are two sources of dispersion. One source, 
due to the discreteness of the wave machine, is encoded in the separation $d$
between the rods. The second source, due to the 
gravitational restoring torque, is encoded in the pendulum-mode frequency. For 
the PASCO wave machines, the dispersion due to the 
discreteness of the rods is negligible, because the wavelength of the 
waves traveling on the machines is always much larger than $d$. Thus, 
$kd$ is always small and we can approximate
$\sin(kd/2)\approx kd/2$. Within this approximation, 
Eq.~(\ref{dispredtod}) becomes
\begin{equation}
       \omega ^2 \approx\omegap ^2 +c^2k^2, 
        \label{dispredtodgonly}
\end{equation}
where
\begin{equation}
       c=\sqrt{\frac{\kappa_L L}{I/L}} \equiv 
            \sqrt{ \frac{\rm stiffness}{\rm inertia}} 
        \label{shivespeed}
\end{equation}
is the wave speed provided by Shive\cite{SHIVE} 
(we provide an alternative derivation of the wave speed in Appendix~\ref{sec:wsam}).  In 
Eq.~(\ref{shivespeed}), $\kappa _L=\kappa_d\, d/L$ is the torsion constant of the 
entire wire of length $L$, and $I = N I_{\rm rod}$ is the moment 
of inertia of the whole set of $N$ rods.

The phase and group velocities can be easily calculated from 
Eq.~(\ref{dispredtodgonly}), giving
\begin{equation}
     v_{\rm phase}=\frac{\omega}{k} \approx\sqrt{\frac{\omegap ^2}{k^2} + c^2} 
 \end{equation}
and
\begin{equation}
     v_{\rm group}=\frac{\rmd \omega}{\rmd k} \approx
        \frac{c^2k}{\sqrt{\omegap ^2 + c^2k^2}} =
         \frac{c^2}{v_{\rm phase}}.
  \end{equation}
Because the phase and group velocities are different, the wave 
machine is dispersive even when we neglect the discreteness of the rods---the
source of dispersion is the gravitational restoring torque.
Interestingly, the gravitationally-induced dispersion disappears if we turn
gravity off.  To see this, we let $g$ go to zero in Eqs.~(\ref{omegap}) 
and~(\ref{dispredtodgonly}) to get
\begin{equation}
       \omega  \approx ck,
        \label{dispredtodgonlyg0}
\end{equation}
which leads to identical expressions for the group and phase velocities:
\begin{equation}
     v_{\rm phase}=\frac{\omega}{k} \approx c,
\end{equation}
\begin{equation}
     v_{\rm group}=\frac{\rmd \omega}{\rmd k} \approx c.
\end{equation}
Thus, when we neglect the discreteness of the rods \textit{and} turn off gravity, 
the wave machine becomes non-dispersive.

Figure~\ref{FIG:generaldis} shows a plot of the dispersion 
relations ($\omega$ vs.~$k$) from Eqs.~(\ref{dispredtod}), (\ref{dispredtodgonly}), and 
(\ref{dispredtodgonlyg0}). From Fig.~\ref{FIG:generaldis}, we can see
that the wave machines have three regions of dispersion. In the first region, 
which corresponds to low wave numbers, the exact dispersion 
relation given by Eq.~(\ref{dispredtod}) is well-approximated 
by Eq.~(\ref{dispredtodgonly}). Thus, for low wave numbers the dispersion is 
due solely to the restoring gravitational torque. In this low-$k$ region, 
we find the pendulum mode (which corresponds to the $y$-intercept) and the 
lowest harmonics of the wave machine. In the second region, which corresponds to
intermediate wave numbers, we have a dispersion-free region, since 
both Eqs.~(\ref{dispredtod}) and~(\ref{dispredtodgonly}) are
well approximated by the linear relationship given by Eq.~(\ref{dispredtodgonlyg0}). There 
is also a third, high-$k$ region of dispersion, where the exact 
dispersion relation [Eq.~(\ref{dispredtod})] is not well-approximated by either
Eq.~(\ref{dispredtodgonly}) or Eq.~(\ref{dispredtodgonlyg0}). The dispersion in
the high-$k$ region is due solely to the discreteness of the rods. In this
high-$k$ region, there is a cut-off frequency, set by the spacing between rods
in the wave machine.\cite{BURGEL} In 
the PASCO wave machines, this high-$k$ region is not attainable.

Another way to see that the dispersion relation at low wave numbers is
due solely to the gravitational torque and not to the discreteness
of the rods is to determine the equation of motion and then find the dispersion relation
when the rods are so close together that they can be considered
a continuum. The resulting dispersion relation is, not surprisingly, given
by Eq.~(\ref{dispredtodgonly}) (see Appendix~\ref{sec:extension} for details). Thus,
even when we ignore the discreteness of the rods from the start,
we still have a gravitationally-induced dispersion at low wave numbers.

We note in passing that the dispersion relation in the wave machine 
[Eq.~(\ref{dispredtodgonly})] is mathematically equivalent to the 
dispersion relation of electromagnetic waves in a plasma at high frequencies
(see Ref.~\onlinecite{JACKSON}). Such dispersion 
in a plasma is enhanced (diminished) whenever the charge of the particles in the 
plasma is increased (decreased), much like the dispersion in a wave machine 
is enhanced (diminished) when the gravitational field strength
is increased (decreased).

\section{Experimental Arrangement}
\label{sec:exparra}

\begin{figure}[!h]
\includegraphics[width=16cm]{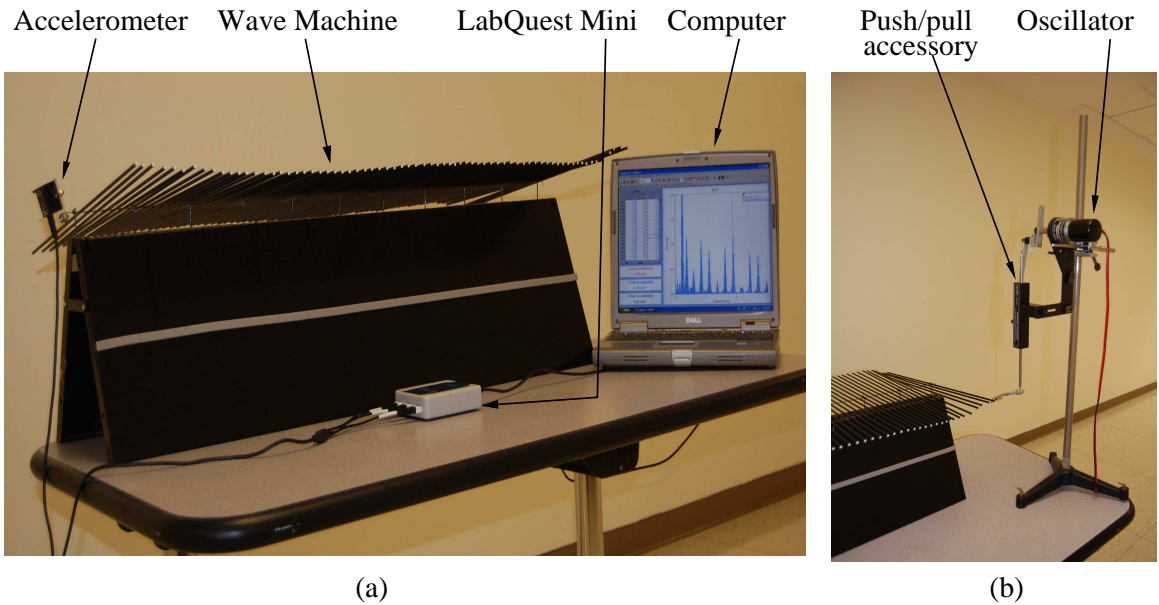}
\caption{(a) Experimental apparatus. (b) The oscillator drives the wave machine 
through PASCO's push/pull accessory, which is attached to one of the
end rods.}
\label{FIG:exap}
\end{figure}

\begin{figure}[!h]
\includegraphics[width=8.5cm]{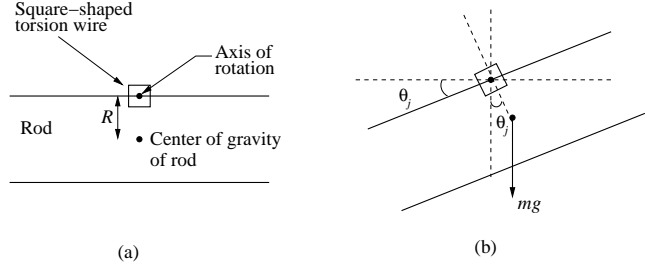}
\caption{Rod in (a) equilibrium position, and (b) tilted by an angle
$\theta _j$. Because the center-of-gravity of the rod is below the axis
of rotation, the weight $mg$ of the rod produces a restoring gravitational
torque of magnitude $mgR\sin (\theta _j)$.}
\label{FIG:torque}
\end{figure}

\begin{figure}[!h]
\includegraphics[width=8.5cm]{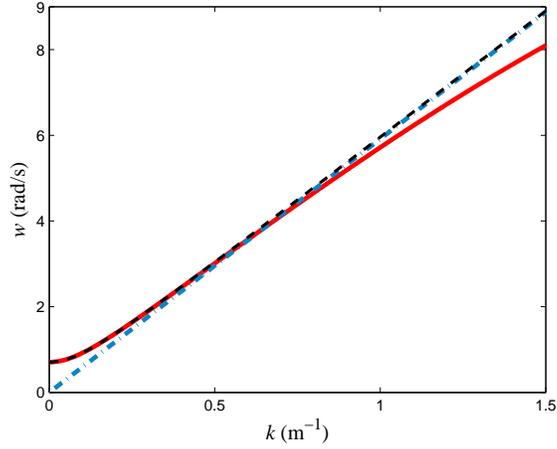}
\caption{Plots of the dispersion relations: Eq.~(\ref{dispredtod}) 
(red, solid line), 
Eq.~(\ref{dispredtodgonly}) (black, vertically-dashed line), and 
Eq.~(\ref{dispredtodgonlyg0}) (blue, dash-dot line). The plots correspond to 
$c=\sqrt{35}$~m/s, $\omegap=0.5$~rad/s, $\omegat = 1$~rad/s, and $d=1$~m.}
\label{FIG:generaldis}
\end{figure}

\begin{figure}[!h]
\includegraphics[width=8.5cm]{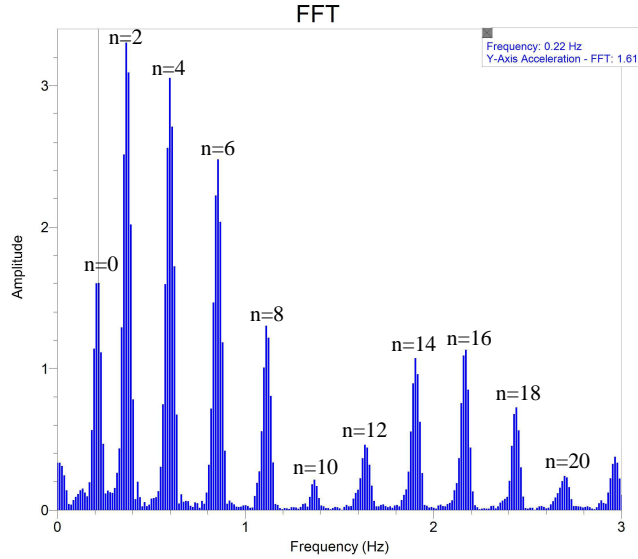}
\caption{FFT frequency spectrum of a wave pulse sent through 
the long-rod wave machine in a one-$g$ environment with free ends.
By clicking on the center of each
peak, the harmonic frequency is displayed in an inset at the top 
right corner. The pendulum mode $(n=0)$ corresponds to the
peak of lowest frequency (0.22\,Hz).}
\label{FIG:FFT}
\end{figure}

An accelerometer (Vernier 3D-BTA) was mounted on one end of the torsion wire 
of the wave machine (see Fig.~\ref{FIG:exap}), and connected to a computer 
through an interface (Vernier LabQuest Mini). A wave pulse was then sent through
the wave machine, and the accelerometer signal was converted into
a frequency spectrum using a built-in Fast Fourier Transform (FFT) routine.
A sample frequency spectrum is shown in Fig.~\ref{FIG:FFT}. By clicking on 
the center of each peak of the frequency spectrum, we can measure the 
frequency of the harmonics excited by the pulse; the resulting frequencies are
listed in Table~\ref{table:freq}. The uncertainty in each frequency
was obtained as the half-width at half-maximum of the peak in the 
frequency spectrum. It should be noted that different pulses may yield slightly
different frequency spectra, and therefore slightly different
harmonic frequencies, especially for the lowest 
harmonics. In such cases, the harmonic frequency can be taken 
as the average of the different measurements.

An alternative way to measure the harmonic frequencies is to excite 
the harmonics of the wave machine individually by way of an
oscillator, as shown in Fig.~\ref{FIG:exap}(b). To accomplish this, we mounted
a push/pull accessory (PASCO ME-8751) on an oscillator (PASCO ME-8750) driven
by a power supply. The push/pull accessory was attached to one of the end rods 
of the wave machine, on the opposite end to the accelerometer. The frequency 
of the oscillator was then increased incrementally until a resonance was obtained.
Near the resonance, the frequency was finely tuned until the amplitude
of the oscillation of the antinodes was as large as possible. The resulting 
frequency spectrum was just a sharp peak centered at the resonant
(harmonic) frequency. By visually inspecting
the number $n$ of antinodes of the harmonic, we were able to determine 
the wave number using a simple standing-wave analysis (see below).
Except for the lowest harmonics, this method resulted in values that
were within 3\% of the values obtained using an FFT analysis of a wave pulse.
For the lowest harmonics ($n\leq 4$), the agreement was about 10\% or better.
The reason for the relatively poor agreement for the lowest harmonics is
that we found it nearly impossible to excite only a single harmonic for
such low frequencies.  Thus, our measured frequencies for the pendulum mode
and the lowest harmonics are the most unreliable.

The main advantage of using an oscillator to measure the frequencies 
is that students can see that each peak of the frequency spectrum 
corresponds to a specific standing wave mode. However, using the 
oscillator has some disadvantages. First, the data acquisition takes 
much longer than using an FFT analysis of a wave pulse. Second, we were 
only able to measure the first ten harmonics when using the oscillator,
whereas the FFT analysis of a wave pulse allowed us to find over nineteen 
harmonics.

The wavelengths of the harmonics produced in a wave machine mimic exactly
the wavelengths of the harmonics produced in a pipe.\cite{ZEMANSKY} When both
ends of the wave machine are free to oscillate (which corresponds to an open 
pipe), the wavelengths can be written
\begin{equation}
           \lambda _n= \frac{4L}{n} \, , \qquad n=2, 4, 6, \ldots
                \quad \text{[\small{two free ends}],}
\end{equation}
where $n$ is the number of antinodes of the wave machine. When 
we fix one end of the wave machine (which corresponds to a 
stopped pipe), the wavelengths can be written
\begin{equation}
           \lambda _n= \frac{4L}{n} \, , \qquad n=1,3, 5,  \ldots
            \quad \text{[{\small single fixed end}].}
\end{equation}
Thus, the wave numbers are given by\cite{TEXT}
\begin{equation}
        k_n= n \frac{\pi}{2L} \, , \qquad n=2, 4, 6, \ldots 
          \quad \text{[{\small two free ends}],}
             \label{wnfe}      
\end{equation}
and
\begin{equation}    
       k_n= n \frac{\pi}{2L} \, , \qquad n=1,3, 5,  \ldots
          \quad \text{[\small{single fixed end}].}
                \label{wnfxe}
\end{equation}

\begin{figure}[!h]
\includegraphics[width=8.5cm]{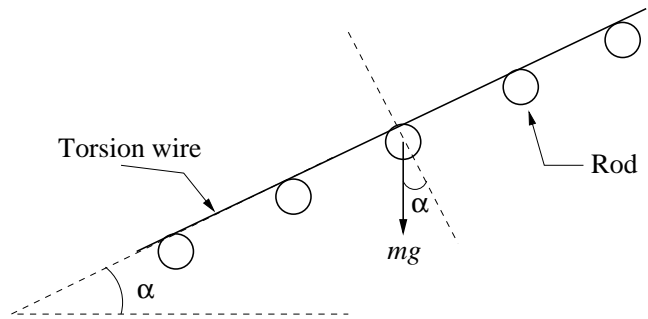}
\caption{Wave machine tilted by an angle $\alpha$. The component of each
rod's weight parallel to the torsion wire ($mg\sin \alpha$) does not
produce any torque. Only the component of the rod's weight perpendicular to the
torsion wire ($mg\cos \alpha$) produces a torque. Hence, tilting the wave
machine by an angle $\alpha$ effectively reduces the strength of gravity to 
$g\cos \alpha$.}
\label{FIG:tiltbyalpha}
\end{figure}

In order to show the dependence of dispersion on gravity, we placed the
wave machines in three different orientations. In the first arrangement the
wave machines are positioned horizontally, so the gravitational field strength
$g$ that appears in Eqs.~(\ref{dispredtod}) and~(\ref{dispredtodgonly}) is
given by 9.8\,m/s$^2$ (one ``$g$''). In the 
second arrangement we tilt the wave machines by an
angle $\alpha \simeq 60^\circ$, thereby reducing the effect of
gravity by $\cos(60^\circ)=1/2$ (see Fig.~\ref{FIG:tiltbyalpha}).
In this arrangement, we thus have a gravitational field strength
of 4.90\,m/s (one-half $g$).
Finally, in the third arrangement we hang the wave machines vertically, 
thereby eliminating the effect of gravity and producing a zero-$g$ environment.
The measured frequencies for both wave machines in these three
arrangements are listed in Table~\ref{table:freq}.

It is important to note that because the wave machines are built to run
horizontally, the friction between the rods and the supports increases
quite a bit when the machines are tilted or hung.
In order to minimize such friction, we connected the torsion 
wire of the wave machine to a fixed pole while the frame of the wave 
machine was immobilized by two supports.\cite{VIDEO}

\section{Comparison between theory and experiment}
\label{sec:expres}

In order to compare theory and experiment, we first need to determine
the wave speeds and pendulum-mode frequencies.  To determine these
values, we perform linear regressions of the $\omega ^2$ vs.~$k^2$
data of Table~\ref{table:freq} using Eq.~(\ref{dispredtodgonly})
for the one-$g$ and the half-$g$ experiments. These regressions yield the 
wave speeds $c$ (the square root of the slope) and the pendulum-mode 
frequencies $\omegap$ (the square root of the 
$y$-intercept). We also perform linear regressions of the
zero-$g$ data of Table~\ref{table:freq} using Eq.~(\ref{dispredtodgonlyg0}). All of
the fits had $R^2=0.9994$ or higher and are summarized in Table II.  From this table, we see 
that the wave speed does not change 
appreciably as the effective gravitational field strength decreases, whereas
the pendulum-mode frequency is seen to decrease. 

In Figs.~6--8 we plot the experimental data from Table~\ref{table:freq}
(one $g$, one-half $g$, and zero $g$, respectively), along with
the theoretical predictions given by either Eq.~(\ref{dispredtodgonly})
or Eq.~(\ref{dispredtodgonlyg0}).  These graphs include only
the data from the long-rod wave machine, since the plots for the
short-rod wave machine are almost identical (except that the frequencies
of the short-rod wave machine are larger than those of the
long-rod wave machine by a factor of about $\sqrt{8}$).
In the graphs, $k$ is plotted in units
of $\pi /2L$, so the wave numbers of the fixed-end harmonics are given
by the odd integers (squares), whereas the wave numbers of the free-end harmonics
are given by even integers (circles). The error bars are about the size of
the data points and have not been included.

\subsection{Reducing the effect of gravity}

\begin{figure}[!h]
\includegraphics[width=8.5cm]{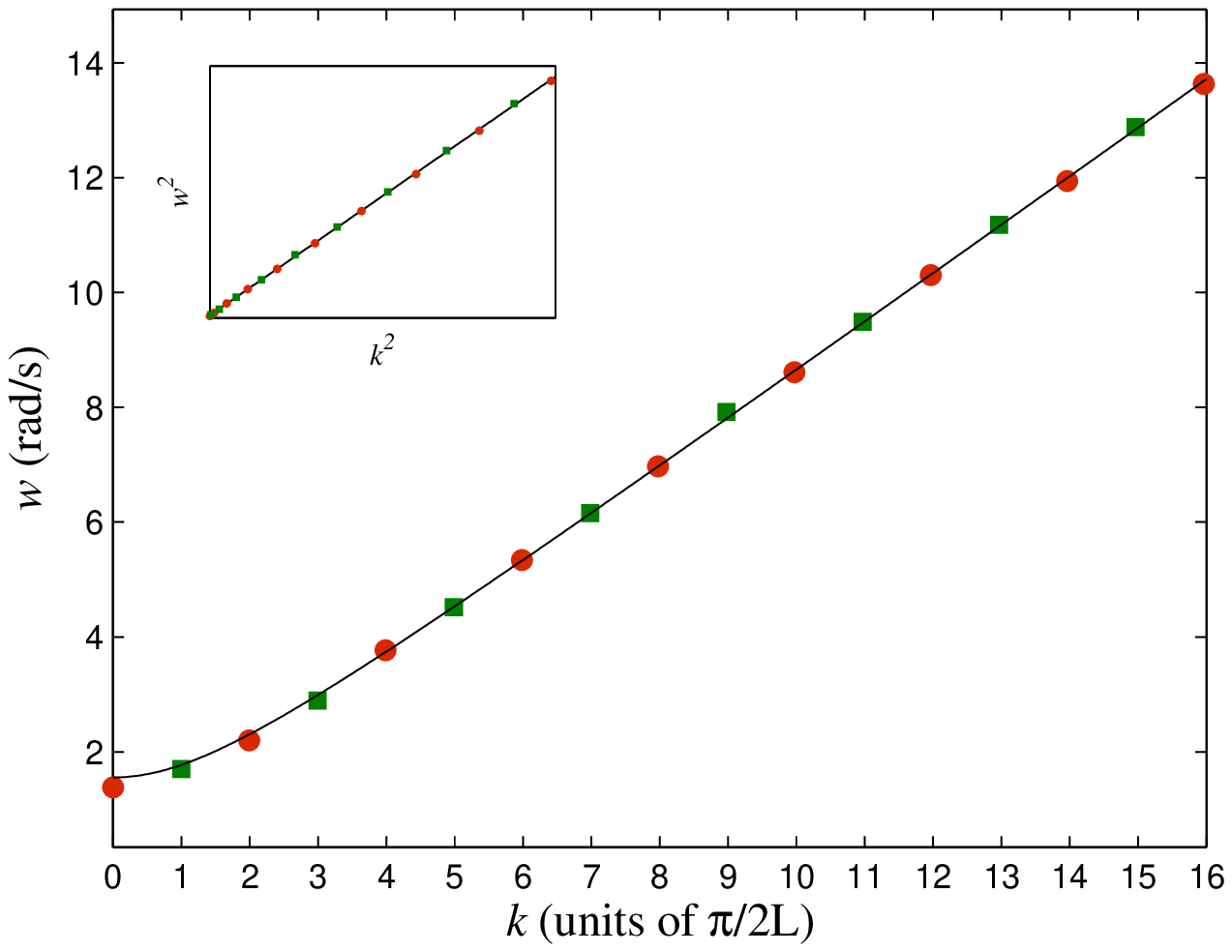}
\caption{Harmonics of the long-rod wave machine in a one-$g$ environment, and 
comparison with the dispersion relation given by 
Eq.~(\ref{dispredtodgonly}) (solid curve).
The red circles represent the free-end harmonics and the green
squares represent the fixed-end harmonics. The inset displays the
linear regression of the $\omega ^2$ vs.~$k^2$ data points.}
\label{FIG:l1g}
\end{figure}

Figure~\ref{FIG:l1g} shows the data for the one-$g$ experiment.
Here we plot the measured harmonic 
frequencies along with the theoretical dispersion 
relation [Eq.~(\ref{dispredtodgonly})] using the values of $c$ and $\omegap$ obtained 
from the linear regression.  The inset shows
the linear fit of the $\omega ^2$ vs.~$k^2$ data points. The agreement 
between theory and experiment shows that
Eq.~(\ref{dispredtodgonly}) fully accounts for the dispersion in the 
wave machines. Thus, for waves that can actually 
be created in the wave machines, the discreteness of the
rods has no observable effect on the dispersion.

\begin{figure}[!h]
\includegraphics[width=8.5cm]{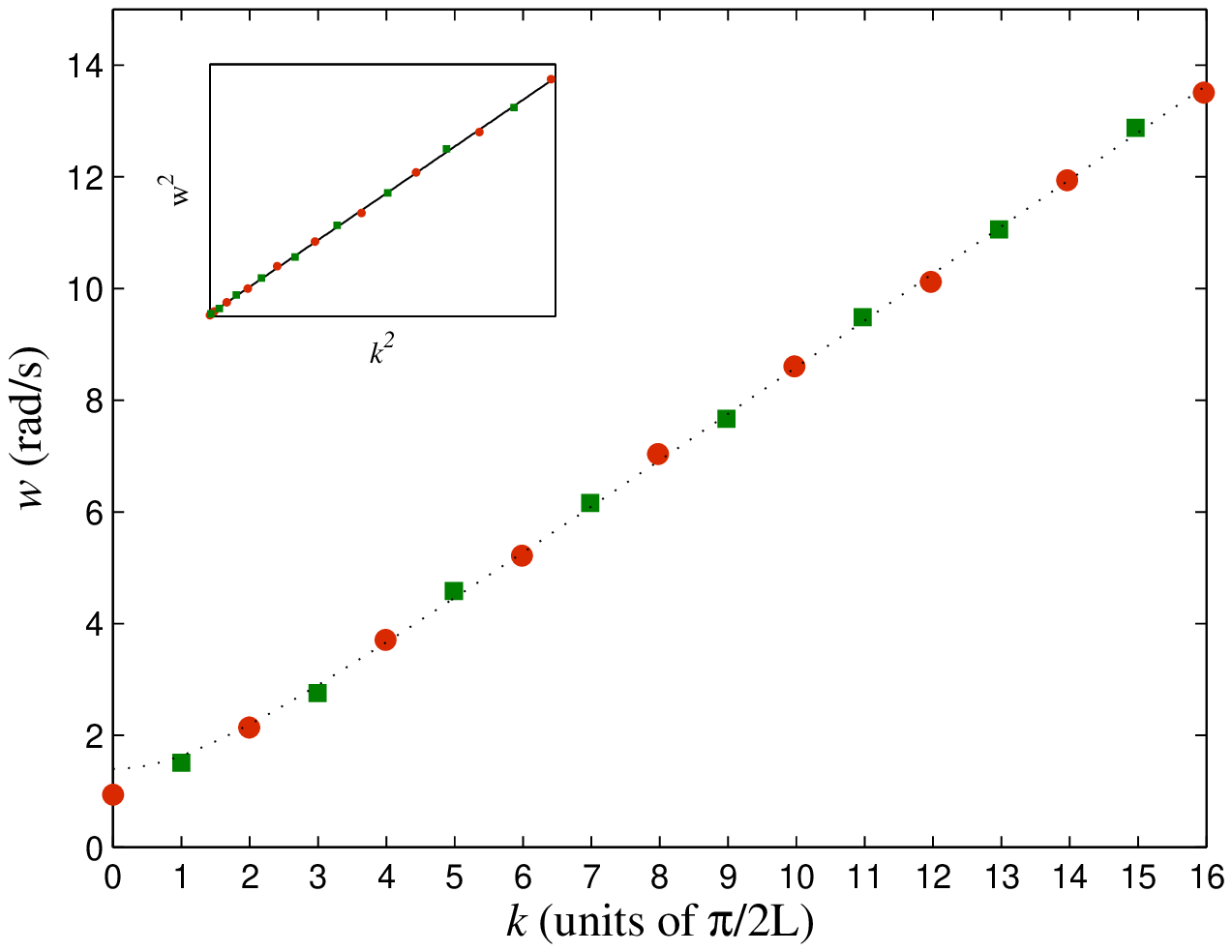}
\caption{Harmonics of the long-rod wave machine in a half-$g$ environment, and 
comparison with the dispersion relation given by Eq.~(\ref{dispredtodgonly}) (dotted curve).
The red circles represent the free-end harmonics, and the green
squares represent the fixed-end harmonics. The inset displays the
linear regression of the $\omega ^2$ vs.~$k^2$ data points.}
\label{FIG:l05g}
\end{figure}

Figure~\ref{FIG:l05g} shows the data for the half-$g$ experiment.
Again we plot the measured harmonic frequencies 
along with the dispersion relation given in Eq.~(\ref{dispredtodgonly}). As
in the one-$g$ experiment, the agreement between theory
and experiment is reasonably good.

\begin{figure}[!h]
\includegraphics[width=8.5cm]{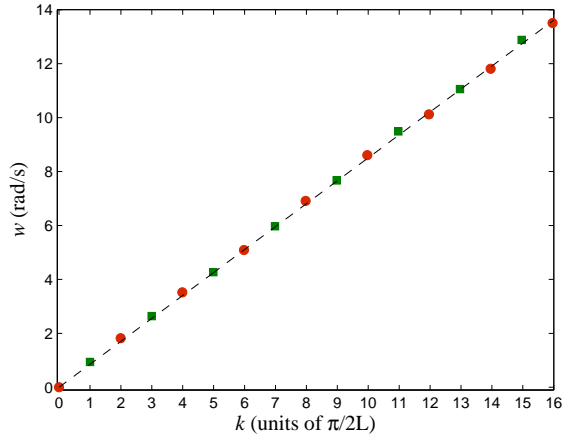}
\caption{Harmonics of the long-rod wave machine in a zero-$g$ environment, and 
comparison with the dispersion relation given by Eq.~(\ref{dispredtodgonlyg0}) (dashed line).
The red circles represent the free-end harmonics, and the green
squares represent the fixed-end harmonics.}
\label{FIG:l0g}
\end{figure}

Figure~\ref{FIG:l0g} shows the data for the zero-$g$ experiment, where
we plot the measured harmonic frequencies along with the 
dispersion relation of Eq.~(\ref{dispredtodgonlyg0}).  The agreement
between experiment and theory shows that the wave machines
become non-dispersive in zero gravity.

\begin{figure}[!h]
\includegraphics[width=8.5cm]{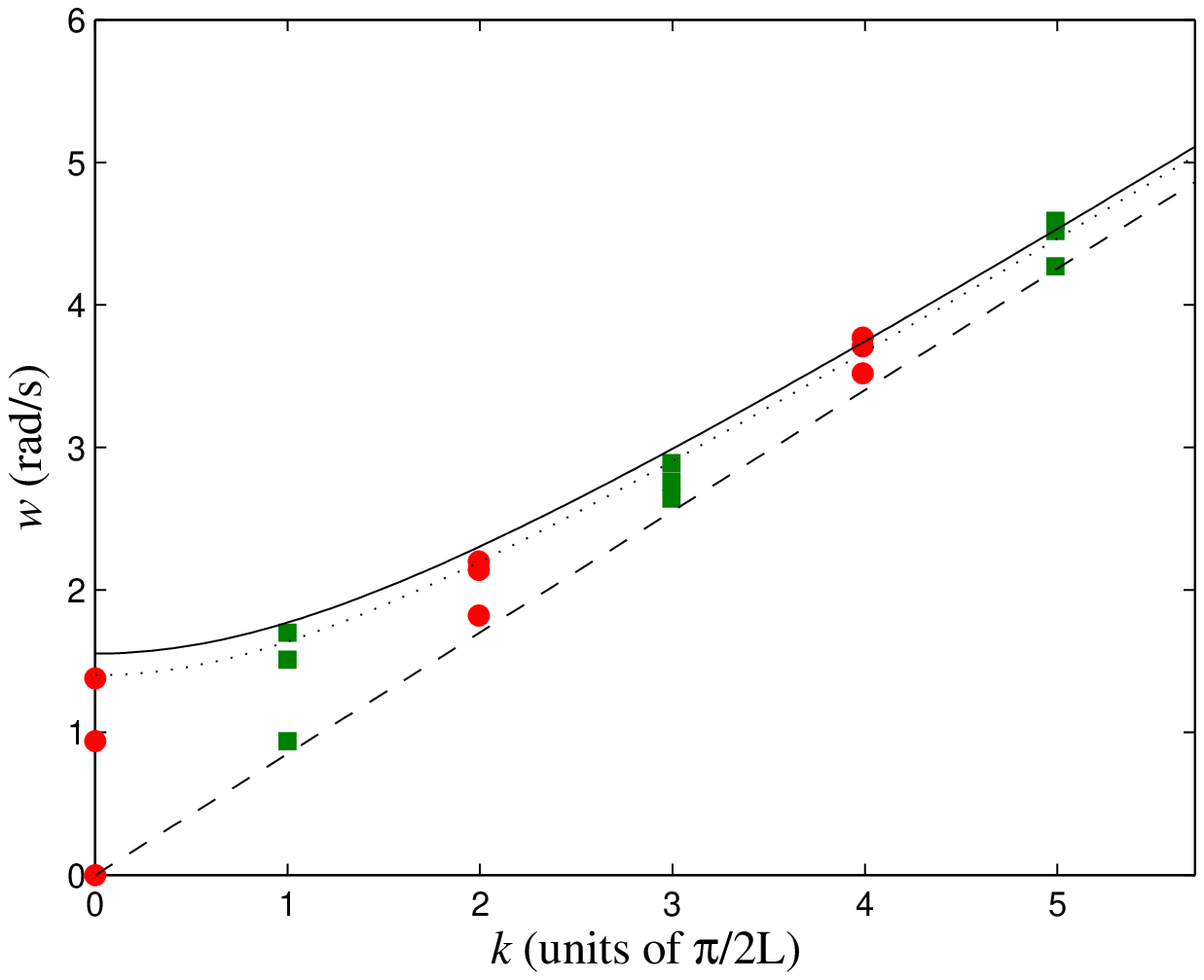}
\caption{Dependence of the dispersion on gravity in the long-rod
wave machine in the low-$k$ region. The solid curve corresponds to one $g$, 
the dotted curve corresponds to one-half $g$, and the dashed line corresponds to zero-$g$.}
\label{FIG:gdl}
\end{figure}

In order to demonstrate more clearly how the gravitational dispersion in the wave 
machine decreases as we reduce the effect of gravity, 
Fig.~\ref{FIG:gdl} shows the pendulum mode and the first five harmonics of the 
long-rod wave machine for all three experiments. Here, we can see how the
harmonic frequencies approach the straight line as gravity is reduced to zero.

\subsection{Ratio of wave speeds}

In the non-dispersive region, the theoretical wave speed 
in a torsional wave machine is given by Eq.~(\ref{shivespeed}). It is easy to
show that because the long-machine rods are twice as long
and twice as massive as the short-machine rods,
and since their moments of inertia are $(1/12) m l^2$, where $m$ is 
the mass of a rod and $l$ its length,\cite{MI} the speeds in the 
short and long wave machines are (theoretically) related by\cite{EL}
\begin{equation}
           \frac{c_{\rm short}}{c_{\rm long}}= \sqrt{8}=2.828 \, .
           \label{wspthe}
\end{equation}
Experimentally, we find that the ratio of the averages of the speeds is
\begin{equation}
           \frac{c_{\rm short}}{c_{\rm long}}= 2.87 \pm 0.27\, ,
\end{equation}
in good agreement with the theoretical result. It should be noted that this
ratio is often reported simply as three.\cite{HARVARD}

\section{Conclusions}
\label{sec:conclus}

We have shown that dispersion in a torsional wave machine is due
to gravity, and that such dispersion diminishes as gravity decreases.
Experimentally, we reduced the effective gravitational strength by tilting
the wave machine. In the limit when gravity goes to zero (in our case, when the wave machines are
hung vertically), we showed that the gravitationally-induced dispersion disappears and the
wave machines becomes non-dispersive. 

Our conclusions are restricted to the PASCO torsional wave machines
used in our experiments (or other similar devices), and do not apply
to all wave machines. For example, there is no gravitational dispersion
in the wave machine used in Refs.~\onlinecite{SKELDON1} and
\onlinecite{SKELDON2} because this wave machine hangs from 
both ends rather than laying on supports.  Thus, the dispersion in this machine
results only from the discreteness of the rods. It should also be noted that in torsional
wave machines of the type used in our experiment, the
dispersion induced by the discreteness of the rods would be measurable
if the rods were separated by a larger distance.

Finally, we would like to note that the two wave machines discussed here
were flown in NASA's reduced gravity flight program in July 2012. The NASA flight allowed 
us to test the theory both at zero\,$g$ and at 1.8\,$g$. In the NASA flight, we 
excited the harmonics of the wave machine using an oscillator that
was turned on and off for brief periods of time. The frequency 
of the oscillator was low, in order to
excite mainly the lowest harmonics. When the oscillator was on, we had free-end
boundary conditions, and when it was off, we had fixed-end boundary 
conditions.

Unfortunately, although our ground test of the experiment worked well, 
the results of the NASA flight do not agree conclusively with the 
results of the present paper.  One possible reason is that, in the experiment 
in the NASA flight, the intervals during which the oscillator was turned 
on and off were not in sync with the periods of climbing (1.8\,$g$) and 
descending (zero\,$g$) of the aircraft.  A better procedure for this 
experiment would be the following: (1) manually excite a wave pulse 
at the beginning of a descending period of the aircraft and collect the
FFT in zero\,$g$; (2) stop the motion of the wave machine
at the end of the descending period; (3) excite a new wave pulse at the
beginning of the climbing period and collect a new FFT in 1.8\,$g$; (4) repeat 
steps (1)--(3) for successive cycles of zero\,$g$ and 1.8\,$g$; (5) after the
flight, obtain the harmonics in 1.8\,$g$ and zero\,$g$ from the FFTs.

\begin{acknowledgements}
The authors would like to thank NASA for selecting the wave-machine 
experiment for its 2012 campaign of reduced gravity flights. The
authors would also like to thank the LUV team (Nicholas Allen, Kirk Goza,
Zach Jones, Jessica Plaia, Aleiya Samad, Aaron Weatherford, and Jacob Wright)
for participating in the NASA flight. Special thanks are due to 
Jim Jordan and the Earth and Space Sciences Department of Lamar University 
for their technical support. The authors are also grateful to the journal's 
reviewers for their suggestions.
One of the authors (RM) acknowledges financial support from 
Ministerio de Ciencia e Innovaci\'on of Spain under 
project TEC2011-24492.
\end{acknowledgements}

\appendix
\section{A continuum torsional wave machine}
\label{sec:extension}

In this appendix, we derive the equation of motion that
describes the motion of a torsional wave machine in the continuum limit that
the distance $d$ between the rods goes to zero. Using this equation of 
motion, we then derive the dispersion relation. 

In a wave machine where the rods are packed together, let us
consider a segment of length $\Delta x$. For small oscillations, Newton's 
second law of rotational motion for such a segment can be written as
\begin{equation}
  I_{\Delta x} \left( \frac{\partial ^2 \theta }{\partial t^2} \right) =
       -m_{\Delta x} g R \theta + 
          \kappa _{\Delta x} (\theta _{x-\Delta x}-\theta) +
           \kappa _{\Delta x} (\theta _{x+\Delta x}-\theta) \, ,
       \label{eqmonrodcon}                          
\end{equation}
where $I_{\Delta x}$ is the moment of inertia of a segment of rods of
length $\Delta x$; $m_{\Delta x}$ is the mass of a segment of rods of length
$\Delta x$; $\theta$, $\theta _{x+\Delta x}$, and $\theta _{x-\Delta x}$
are the angular displacements from equilibrium at positions $x$, 
${x+\Delta x}$, and ${x-\Delta x}$, respectively; 
$\kappa _{\Delta x}$ is the torsion constant of a section of wire of
length $\Delta x$;
and $R$ is the radius of a rod. If the rods are uniformly distributed, the
mass of rods per unit of length is constant, and therefore
\begin{equation}
          \frac{M}{L}=\frac{m_{\Delta x}}{\Delta x} \, ,
       \label{umoinfd}
\end{equation}
where $M$ is the mass of the whole set of rods, and $L$ is the length
of the wave machine. Similarly, if the rods are uniformly distributed, the
moment of inertia per unit length is constant, and therefore
\begin{equation}
          \frac{I}{L}=\frac{I_{\Delta x}}{\Delta x} \, ,
       \label{umoinf}
\end{equation}
where $I$ is the total moment of inertia of the whole set of rods. In addition,
we have that
\begin{equation}
          \kappa _{\Delta x}\Delta x = \kappa _{L}L \, .
       \label{umoinf22}
\end{equation}

Substituting Eqs.~(\ref{umoinfd})--(\ref{umoinf22}) into 
Eq.~(\ref{eqmonrodcon}), and then
multiplying and dividing the second term on the
right-hand-side of Eq.~(\ref{eqmonrodcon}) by $\Delta x$, we obtain
\begin{equation}
  \Delta x \frac{I}{L} \left( \frac{\partial ^2 \theta }{\partial t^2} \right) =
       -\Delta x \frac{M}{L} g R \theta + 
          \kappa _{L} L \left( 
            \frac{\theta _{x+\Delta x}-\theta }{\Delta x} +
        \frac{\theta _{x-\Delta x}-\theta}{\Delta x} \right) \, .
       \label{eqmonrodcon2}                          
\end{equation}
In the limit that $\Delta x$ is very small, we find that
\begin{equation}
  \Delta x \frac{I}{L} \left( \frac{\partial ^2 \theta }{\partial t^2} \right) =
       -\Delta x \frac{M}{L} g R \theta + 
          \kappa _{L}L \left[ \theta '(x) -\theta '(x-\Delta x) \right]
                   \, .
       \label{eqmonrodcon3}                          
\end{equation}
Again, taking the limit that $\Delta x\rightarrow 0$, we arrive at the
equation of motion
\begin{equation}
  \frac{I}{L} \left( \frac{\partial ^2 \theta }{\partial t^2} \right) =
       - \frac{M}{L} g R \theta + 
          \kappa _{L}L \frac{\partial ^2 \theta }{\partial x^2} 
                   \, .
       \label{eqmonrodcon5}                          
\end{equation}
By substituting a fixed-amplitude sinusoidal wave 
$\theta (t)=A e^{\rmi (kx-\omega t)}$ into Eq.~(\ref{eqmonrodcon5}),
we get the dispersion relation given by Eq.~(\ref{dispredtodgonly}). Thus, 
not surprisingly, when we neglect the discreteness of the rods from
the beginning, we find the same dispersion relation as was found
when the discreteness of the rods was assumed to be negligible.

\section{Wave speed}
\label{sec:wsam}

In Ref.~\onlinecite{SHIVE}, Shive states that the wave speed in a wave machine 
is given by Eq.~(\ref{shivespeed}). Such a wave speed can be derived
by, for example, neglecting the restoring gravitational torque in 
Eq.~(\ref{eqmonrodcon5}).  Here, we provide an alternative derivation. Such a
derivation can be viewed as the rotational analog of the derivation 
of the wave speed on a string (see, for example, Ref.~\onlinecite{ZEMANSKY}, page~499). 

Let us consider the continuum version of the wave machine, neglecting
the effect of the restoring gravitational torque. For this, let us 
imagine that we exert a torque $\tau$ on one of the ends of the wave 
machine. This torque will produce a disturbance that propagates with speed
$c$. After a (small) time $\Delta t$, the disturbance has traveled a 
(small) distance 
$\Delta x= c \Delta t$ along the wave machine. The torque exerted on the 
wave machine is given by 
\begin{equation}
             \tau =\kappa _{\Delta x} \Delta \theta \, .
\end{equation}
Due to the torque, the angular momentum
of a segment of the wave machine of length $\Delta x$ will change 
from 0 to $\ell$. On the one hand, ${\ell}$ is given by
\begin{equation}
         \ell = \tau \Delta t=  \kappa _{\Delta x} \Delta \theta \Delta t,
            \label{Ltau}
\end{equation}
and on the other it is given by
\begin{equation}
        \ell = 
           I_{\Delta x}\omega = I_{\Delta x}\frac{\Delta \theta}{\Delta t}.
       \label{LIW}
\end{equation}
Combining Eqs.~(\ref{Ltau}) and (\ref{LIW}) gives
\begin{equation}
       I_{\Delta x}\frac{\Delta \theta}{\Delta t} = 
\kappa _{\Delta x} {\Delta \theta}{\Delta t} \, ,
        \label{interd}
\end{equation}
and by canceling $\Delta \theta$ and multiplying by $\Delta x$ we get
\begin{equation}
       I_{\Delta x}\frac{\Delta x}{\Delta t} = 
\kappa _{\Delta x} \Delta x {\Delta t} \, .
\end{equation}
Because the speed of the wave is $c=\Delta x/\Delta t$, and using
$\kappa _{\Delta x} \Delta x =\kappa _{L}L$, we end up with
\begin{equation}
       I_{\Delta x}c = 
\kappa _{L} L {\Delta t} \, .
\end{equation}
Since we are assuming that the wave machine is uniform, we have that 
$I_{\Delta x}/\Delta x= I/L$ and therefore
\begin{equation}
       \Delta x \, (I/L)\, c = 
\kappa _{L} L {\Delta t} \, .
\end{equation}
Cross multiplication then yields
\begin{equation}
       \frac{\Delta x}{\Delta t} c = 
                     \frac{\kappa _{L} L}{I/L},
\end{equation}
or
\begin{equation}
          c^2 = \frac{\kappa _{L} L}{I/L} \, ,
\end{equation}
which coincides with Eq.~(\ref{shivespeed}).

\begin{table}[!h]
\centering
\caption{Frequencies and wave numbers of all the harmonics. 
All frequencies are in Hz and the error for each experiment
is given in parentheses.
Equations~(\ref{wnfe}) and~(\ref{wnfxe}) were used to calculate the wave
numbers; the length $L$ of the long-rod and short-rod wave machines
are, respectively, $0.890 \pm 0.004$\,m and $0.900 \pm 0.004$\,m (the
uncertainty being equal to the diameter of one rod).} 
\vspace{6pt}
\begin{tabular}{c c| c c c | c c c} 
\hline\hline
\multicolumn{2}{c}{\quad} & \multicolumn{6}{c}{Harmonic frequencies (Hz)} \\
\cline{3-8}
\multicolumn{1}{c}{\quad}&
\multicolumn{1}{c}{\quad}& \multicolumn{3}{c}{Long-rod  machine} 
& \multicolumn{3}{c}{Short-rod  machine} 
\\
\hline
 $n$     &  $k_n$  &   one $g$   &  one-half $g$   & 
 zero $g$   
&  one $g$   &  one-half $g$   &  zero $g$   \\ [-8pt]
  &   &   ($\pm0.03$)   &  ($\pm0.05$)   & ($\pm0.08$)   
&  ($\pm0.1$)   &  ($\pm0.4$)   &  ($\pm0.7$)   \\ [0.5ex]

\hline 
0 & 0                & 0.22  & 0.15  & 0  &  0.44  & 0.29   & 0     \\ 
1 & $\frac{\pi}{2L}$ & 0.27  & 0.24  & 0.15  & 0.59   & 0.54   & 0.44  \\
2 & $\frac{\pi}{L}$ & 0.35  & 0.34  & 0.29  & 0.93   & 0.88   & 0.83  \\
3 & $\frac{3\pi}{2L}$ & 0.46  & 0.44  & 0.42  & 1.17   & 1.22   & 1.17  \\
4 & $\frac{2\pi}{L}$ & 0.60  & 0.59  & 0.56  & 1.66   & 1.66   & 1.56  \\ 
5 & $\frac{5\pi}{2L}$ & 0.72  & 0.73  & 0.68  & 1.95   & 1.98   & 1.96  \\  
6 & $\frac{3\pi}{L}$ & 0.85  & 0.83  & 0.81  & 2.39   & 2.39   & 2.34  \\ 
7 & $\frac{7\pi}{2L}$ & 0.98  & 0.98  & 0.95  & 2.73   & 2.73   & 2.73  \\
8 & $\frac{4\pi}{L}$ & 1.11  & 1.12  & 1.10  & 3.17   & 3.13   & 3.11  \\
9 & $\frac{9\pi}{2L}$ & 1.26  & 1.22  & 1.22  & 3.52   & 3.52   & 3.52  \\
10 & $\frac{5\pi}{L}$ & 1.37 & 1.37  & 1.37  & 3.91  & 3.91   & 3.83  \\ 
11 & $\frac{11\pi}{2L}$ & 1.51 & 1.51  & 1.51  & 4.30  & 4.25   & 4.25  \\
12& $\frac{6\pi}{L}$ & 1.64 & 1.61  & 1.61  & 4.64  & 4.67   & 4.61  \\
13 & $\frac{13\pi}{2L}$ & 1.78 & 1.76  & 1.76  & 5.08  & 5.08   & 5.08  \\
14 & $\frac{7\pi}{L}$ & 1.90 & 1.90  & 1.88  & 5.42  & 5.47   & 5.42  \\ 
15 & $\frac{15\pi}{2L}$ & 2.05 & 2.05  & 2.05  & 5.86  & 5.76   & 5.86  \\ 
16 & $\frac{8\pi}{L}$ & 2.17 & 2.15  & 2.15  & 6.20  & 6.20   & 6.20  \\
17 & $\frac{17\pi}{2L}$ & 2.32 & 2.29  & 2.32  & 6.64  & 6.59   & 6.54  \\ 
18 & $\frac{9\pi}{L}$ & 2.44 & 2.44  & 2.42  & 6.93  & 6.98   & 6.88  \\ [1ex]
\hline\hline
\end{tabular}
\label{table:freq}
\end{table}

\newpage

\begin{table}[h!] 
\caption{Results of linear regressions of the Table~\ref{table:freq} data
as described in the text.  All  fits had $R^2=0.9994$ or higher.}
\vspace{6pt}
\centering
\begin{tabular}{c| c c | c c} 
\hline\hline
\multicolumn{1}{c}{\quad}& \multicolumn{2}{c}{Long-rod wave machine} 
& \multicolumn{2}{c}{Short-rod wave machine} 
\\
\hline
\quad Effective $g$ \quad \quad   &\quad  $c$ (m/s) \quad   & 
\quad $\omega_{\rm p}$ (rad/s) \quad \quad & 
 \quad $c$ (m/s) \quad    & \quad $\omega_{\rm p}$ (rad/s) \quad \quad \\ [0.5ex]
\hline 
one $g$   & \ 0.4867 $\pm$ 0.0007   & 1.6 $\pm$ 0.1  & \ 1.393 $\pm$ 0.003  &
                        3.1 $\pm$ 0.6           \\ 
one-half $g$ & \ 0.484 $\pm$ 0.001   & 1.4 $\pm$ 0.2  & \ 1.392 $\pm$ 0.002   & 
                        2.9 $\pm$ 0.5   \\ 
zero $g$  & \ 0.479 $\pm$ 0.002   & 0.08 $\pm$ 0.04   & \ 1.382 $\pm$ 0.005  &
                    0.23 $\pm$ 0.09    \\ 
\hline \hline
\end{tabular}
\end{table}

\vskip5cm

\newpage

\vskip4cm

\newpage

\vskip3cm

\newpage

\vskip2cm

\newpage

\vskip2cm

\begin{thebibliography}{99}
\bibitem{TEXT} It should be noted
that most textbooks denote the free-end wavelengths by 
$\lambda_n=2L/n$, with $n=1,2,3, \dots$, and the (single) fixed-end wavelengths 
by $\lambda_n =4L/n$, with $n=1,3,5,\ldots$\,. We instead choose a quarter
wavelength as the basic unit of length, so the free-end
harmonics are associated with even integers, as
in Eq.~(\ref{wnfe}), whereas the (single) fixed-end
harmonics are associated with odd integers, as in
Eq.~(\ref{wnfxe}). Such a choice shows more clearly how the
free-end and fixed-end harmonics alternate.
\bibitem{MI} The moment of inertia of each rod is actually
$(1/12) m l^2 + (5/4)mR^2$, where $R$ is the radius of the
rod. However, because $R\ll l$, the moment
of inertia is well approximated by $(1/12) m l^2$.
\bibitem{EL} For simplicity, we have assumed that 
the lengths of the short and the long wave machines are the same, 
since taking into account the
actual lengths (0.89\,m and 0.90\,m) barely changes the 
ratio given in Eq.~(\ref{wspthe}).
\end{thebibliography}
\end{document}